\documentclass{article}

\usepackage{amssymb,amsmath,bm,graphicx,multirow}
\usepackage{cite}

\def\rd{\mathrm{d}}

\begin{document}
\begin{titlepage}
\begin{center}
\noindent{\large\textbf{New identities for sessile drops}}
\vspace{2\baselineskip}

Maryam Hajirahimi~\footnote{Email: m\_hajirahimi@azad.ac.ir}$^{~(1}$\\
Fahimeh Mokhtari$^{~(2}$\\
Amir H. Fatollahi~\footnote{Email: fath@alzahra.ac.ir}$^{~(2}$
\vspace{0.7cm}

\textit{1)  Physics Group, South Tehran Branch, Islamic Azad University, \\ P. O. Box 11365, Tehran 4435, Iran,
 \\
\vspace{0.5cm}
2)  Department of Physics, Alzahra University, \\ P. O. Box  19938, Tehran  91167, Iran
}
\end{center}
\vspace{\baselineskip}
\begin{abstract}
A new set of mathematical identities is presented for axi-symmetric sessile
drops on flat and curved substrates.
The geometrical parameters, including the apex curvature and height,
and the contact radius, are related by the identities.
The validity of the identities are checked by various numerical solutions
both for flat and curved substrates.
\end{abstract}

\vspace{1cm}


\end{titlepage}

\section{Introduction}
The study of drops with the effect of gravity being balanced with
the surface tension goes back to more than a century \cite{adams},
followed by renewal updates extending and refining the original
treatment \cite{staicop,padday,hartland}. More recent efforts
have concentrated on presenting approximate analytical solutions
or developing more efficient methods for numerical solutions
\cite{chester,ehrlich,smith,shan82,shan84,rienstra,ryley,lehman,obrien,maze, neumann,kwok,graham,neum97,kwok91,fathscr}.

Heuristically, the balance between the surface effects and the bulk ones would fix the profile of
a drop. While the gravity lowers the center of mass, the surface tension ($\gamma$)
tends to decrease the surface, and the adhesion coefficient ($\sigma$) tends to increase the surface of the contact region. For a drop with volume $V$, density $\varrho$ and comparable
surface effects (\textit{i.e.} $\sigma\sim\gamma$), the so-called Bond number defined by the dimensionless combination $V^{2/3}\varrho\, g/\gamma$ would determine whether weight has the dominant contribution or not.

Mathematically, at every point of the drop's surface the Young-Laplace relation holds,
\begin{equation}\label{1}
  \gamma \,\bigg (\frac{1}{R_1} + \frac{1}{R_2} \bigg)=\Delta p
\end{equation}
\noindent where  $(R_1,R_2)$ are two principal
radii of curvature at the point, and $\Delta p\equiv p_\mathrm{l}-p_\mathrm{v}$ is the
pressure jump across the liquid-vapor interface. As the hydrostatic laws express $\Delta p$ in terms of the surface equation, the Young-Laplace relation is the differential equation by which, together with the boundary conditions, the drop's profile is determined.
As one of the boundary conditions, the contact angle ($\vartheta$) is fixed by the Young equation
\begin{equation}\label{2}
\cos \vartheta =  \frac{\sigma}{\gamma} -1.
\end{equation}

The purpose here is to present a set of mathematical identities for sessile drops.
In particular, for the cases of sessile drops on flat and curved
substrates, by direct integration of Young-Laplace relation over
the entire surface of drop, exact identities are derived. The geometrical
parameters of drop, including the height and
curvature at the apex and the contact radius of the drop
are related by the identities. The importance of the mentioned parameters
is that, they are initially unknown, and are determined only after
the complete solution is available.
The validity of the identities are checked by various numerical solutions
both for flat and curved substrates.


\section{The mathematical setup and derivation}
Using the cylindrical coordinate setup given in Fig.~1 the total curvature of the axi-symmetric surface  $z=f (\rho) $ is given by
\begin{equation}\label{3}
\frac{1}{R_1} + \frac{1}{R_2} = \frac{1}{\rho}~\frac{\rd }{\rd \rho}\bigg(\rho \,
\frac {|f '|} {\sqrt {1 + f '^ {\, 2}}} \bigg),
\end{equation}
\noindent where $ f '=\rd  f/\rd  \rho $. On the other hand, the pressure jump in presence of gravity gets contribution from the weight of the drop's layers as well, leading to
\begin{equation}\label{4}
\Delta p (z) =\Delta p_\gamma +  \varrho g (h - z)
\end{equation}
in which $h$ is the height of the drop's apex, and $\Delta p_\gamma$ is a constant
representing the pressure jump due to the surface tension. So, the Young-Laplace relation reads
\begin{equation}\label{5}
\mp\frac{1}{\rho} \frac{\rd }{\rd  \rho} \bigg(\rho \frac {f '_\pm} {\sqrt {1 + f '^ {\, 2}_\pm}} \bigg) =  2 \kappa +  \frac{\varrho g}{\gamma} (h- f_\pm).
\end{equation}
in which $f_+$ and $f_-$ are denoting the upper and lower parts of the drop, respectively (see Fig.~1b), and $\kappa:=\Delta p_\gamma/(2\,\gamma)$. At apex ($h=f_+(0)$) we have $R_1=R_2$, and so by (\ref{1}), $\kappa$ is simply the curvature at apex.

The main issue with equation (\ref{5}) is that the parameters $h$ and $\kappa$ are not known at the first place, and would be determined only after the complete solution is available. So at starting point the main equation is not fully known. Further, the contact radius $\rho_0$
(Fig.~1), as the limiting value for the variable $\rho$, is not known at first place.
As will be seen shortly, by integrating the Young-Laplace relation an identity is obtained
which relates the three unknown parameters in a very helpful way.
Hereafter, the cases for flat and curved substrates are considered separately.

\begin{figure}[t]
\begin{center}
\includegraphics[width=1.0\columnwidth]{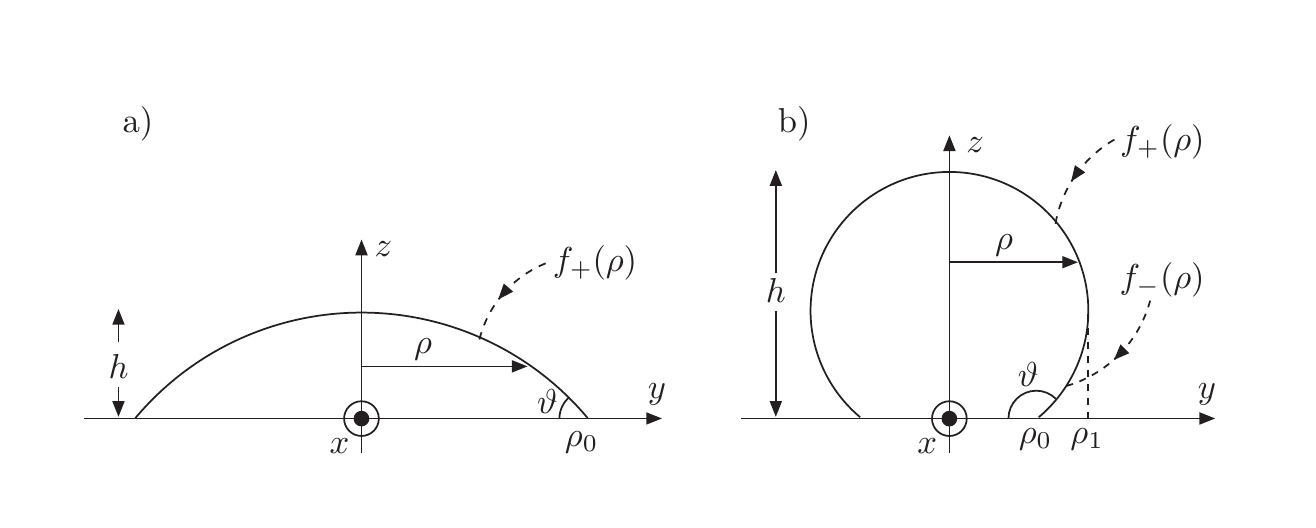}
\caption{\small The geometry of the mathematical setup for: a) $\vartheta < 90^\circ$, b)
$\vartheta > 90^\circ$. }
\end{center}
\end{figure}


\subsection{Flat substrate}
The boundary conditions for $\vartheta>90^\circ$ are:
\begin{align} \label{6}
f'_+(0)&=0\\ \label{7}
f'_-(\rho_0)&=-\tan\vartheta,\\ \label{8}
f_-(\rho_0)&=0.
\end{align}
We mention $f'_+<0$ and $f'_->0$.
In case with $\vartheta<90^\circ$ (\ref{7}) and (\ref{8}) are valid for $f_+$.
In what follows we mainly consider the case with $\vartheta > 90^\circ$. The generalization to case with $\vartheta < 90^\circ$  is rather straightforward. Integrating the Young-Laplace relation for the upper and lower parts of drop leads to
\begin{align}\label {9}
 \rho_1 &=    \left( \kappa +\frac{\varrho g}{2\gamma}h\right) \rho_1 ^ 2 -
\frac {\varrho g} {\gamma} \int_ {0} ^ {\rho_1} \rho f_+(\rho) \rd  \rho \\
\label{10}
\rho_1- \rho_0 \sin \vartheta  & =  \left( \kappa +\frac{\varrho g}{2\gamma}h\right)
(\rho_1 ^ 2 - \rho_0 ^ 2) - \frac {\varrho g} {\gamma} \int_ {\rho_0} ^ {\rho_1} \rho f_-(\rho)\rd  \rho
\end{align}
in which we have used
\begin{equation}\label{11}
\left. \frac  {f _-'} {\sqrt {1 + f_-'^{2}}} \right|_{\rho_0}\!\!\!\! =
\frac {-\tan \vartheta } {\sqrt {1 + \tan ^ 2 \vartheta }} = \sin \vartheta
\end{equation}
for $\vartheta > 90^\circ$. Subtracting (\ref{9}) and (\ref{10}) leads to the identity
\begin{equation}\label{12}
\kappa +\frac{\varrho g}{2\gamma}h
= \frac{\sin \vartheta }{\rho_0} + \frac {\varrho g V} {2 \pi \gamma \,\rho_0^2}
\end{equation}
in which we have used the relation for the volume of drop,
\begin{equation} \label {13}
\frac  {V} {2 \pi}  = \int_0 ^ {\rho_1} \rho f_ +(\rho) \rd  \rho - \int_ {\rho_0} ^ {\rho_1} \rho f_- (\rho)\rd  \rho.
\end{equation}
\noindent  It is easy to show that identity (\ref{12}) is valid for the acute contact angle
($\vartheta < 90^\circ$) as well. It is emphasized in (\ref{12}) no approximation is
used, hence it is an exact relation.

It would be useful to check the above identity for the case with absence of gravity, in
which, as only the surface effects are present, the drop's surface is part of sphere.
By direct insertion it can be seen that the following satisfies the Young-Laplace relation
(\ref{5}),
\begin{equation}\label{14}
z = f_{0\pm} (\rho) = \pm \sqrt {R^2 - \rho ^ 2} + z_0
\end{equation}
\noindent representing a sphere with radius $R $
whose center is located at  $z=z_0$.
Following a simple geometrical argument in the sphere (see Fig.~1),
we have
\begin{equation}\label{15}
\rho_0=R\sin\vartheta,~~~~~~z_0=-R\cos \vartheta.
\end{equation}
As on the surface of a sphere the curvature is constant, we have $\kappa=1/R$.
By setting $g=0$ in the identity (\ref{12}), it is simply satisfied by the given values.


\subsection{Curved substrate}

\begin{figure}[h]
\begin{center}
\includegraphics[width=0.9\columnwidth]{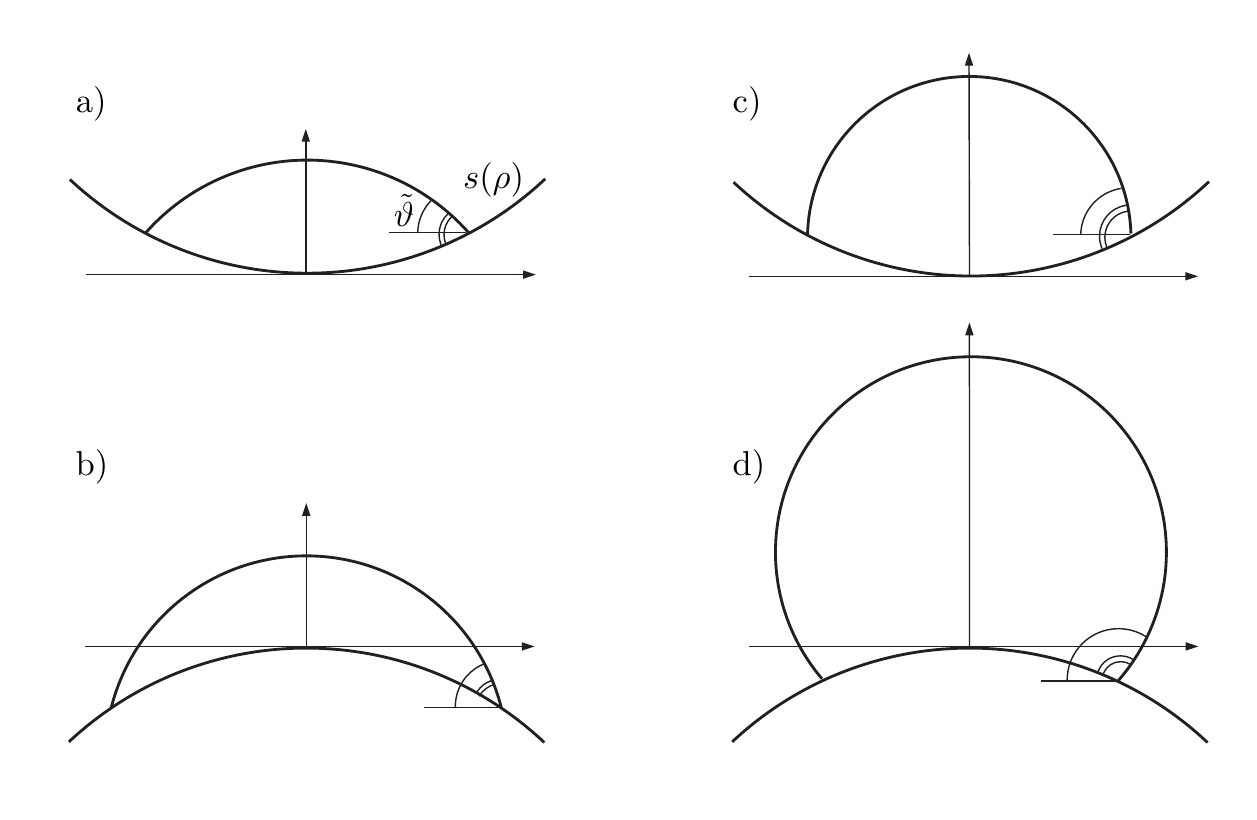}
\caption{\small The four possible situations for drop on curved substrate $s(\rho)$, and definition of
the angle $\tilde \vartheta $ (single-arc), as the angle between the drop and horizontal line at the contact point. The contact angle is represented by double-arc.
Other parameters are the same as Fig.~1. }
\end{center}
\end{figure}

In the lines similar to the case for the flat case, we can derive the identities for curved substrate
as well. The main differences, as follow, appear in the boundary condition, and in
the way that volume of drop comes to the play. The possible situations
are represented in Figs.~2~\&~3.
Here the angle between the drop's surface at contact point and the horizontal
line, denoted by $\tilde\vartheta$ in Figs.~2~\&~3,
appears as part of the boundary conditions.
First let us consider cases shown in Fig.~2. For cases with obtuse angle,
the boundary conditions read:
\begin{align} \label{16}
f'_+(0)&=0\\ \label{17}
f'_-(\rho_0)&=-\tan\tilde\vartheta.
\end{align}
Again, in case with $\vartheta<90^\circ$ (\ref{17}) is valid for $f_+$.
In what follows we mainly consider the case with $\vartheta > 90^\circ$.
The generalization to case with $\vartheta < 90^\circ$  is rather straightforward. Integrating the Young-Laplace relation for the upper and lower parts of drop leads to
\begin{align}\label {18}
 \rho_1 &=    \left( \kappa +\frac{\varrho g}{2\gamma}h\right) \rho_1 ^ 2 -
\frac {\varrho g} {\gamma} \int_ {0} ^ {\rho_1} \rho f_+(\rho) \rd  \rho \\
\label{19}
\rho_1- \rho_0 \sin \tilde\vartheta  & =  \left( \kappa +\frac{\varrho g}{2\gamma}h\right)
(\rho_1 ^ 2 - \rho_0 ^ 2) - \frac {\varrho g} {\gamma} \int_ {\rho_0} ^ {\rho_1} \rho f_-(\rho)\rd  \rho
\end{align}
in which, as mentioned before, $\tilde\vartheta$ is the angle between the drop's surface
at contact point and horizontal line (Fig.~2), for which we have
\begin{equation}\label{20}
\left. \frac  {f _-'} {\sqrt {1 + f_-'^{2}}} \right|_{\rho_0}\!\!\!\! =
\frac {-\tan \tilde\vartheta } {\sqrt {1 + \tan ^ 2 \tilde\vartheta }} = \sin \tilde\vartheta.
\end{equation}
Subtracting (\ref{18}) and (\ref{19}) leads to the identity
\begin{equation}\label{21}
\kappa +\frac{\varrho g}{2\gamma}h
= \frac{\sin \tilde\vartheta }{\rho_0} + \frac {\varrho g} {2 \pi \gamma \,\rho_0^2}
(V+V_s)
\end{equation}
in which we have used the following relations for the volumes,
\begin{align}
\label{22}
\frac  {V} {2 \pi} & = \int_0 ^ {\rho_1} \rho f_ +(\rho) \rd  \rho - \int_ {\rho_0} ^ {\rho_1}
\rho f_- (\rho)\rd  \rho -\frac{V_s}{2\pi},\\
\label{23}
\frac  {V_s} {2 \pi} &= \int_0 ^ {\rho_0} \rho\, s(\rho) \rd  \rho,
\end{align}
\noindent in which $s(\rho)$ is the equation of the substrate (Fig.~2).
It is easy to show that identity (\ref{21}) is valid for the acute contact angle
as well. Again, it is emphasized in (\ref{21}) no approximation is
used, hence it is an exact relation.

\begin{figure}[t]
\begin{center}
\includegraphics[width=0.4\columnwidth]{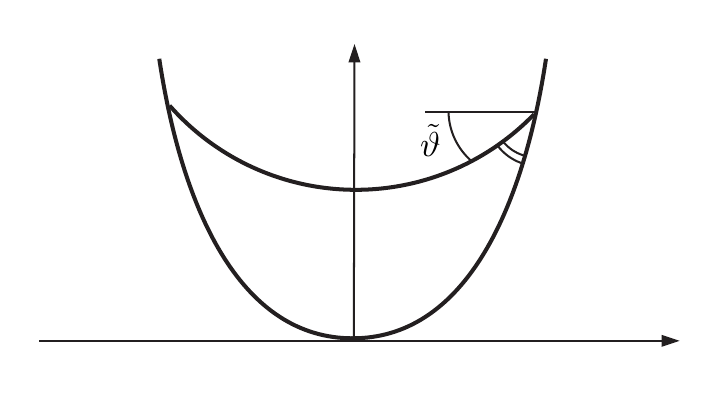}
\caption{\small The case for drop with acute contact angle (double-arc) but concave
surface. }
\end{center}
\end{figure}

Now we come to the case shown in Fig.~3, in which $\rho=0$ makes a
dimple. For this case the Young-Laplace relation reads:
\begin{equation}\label{24}
\frac{1}{\rho} \frac{\rd }{\rd  \rho} \bigg(\rho \frac {f '_-} {\sqrt {1 + f '^ {\, 2}_-}} \bigg) =
2 \kappa -  \frac{\varrho g}{\gamma} (h- f_-),
\end{equation}
by which after integration one finds
\begin{align}\label{25}
\rho_0 \sin \tilde\vartheta  & =  \left( \kappa -\frac{\varrho g}{2\gamma}h\right)
\rho_0 ^ 2 + \frac {\varrho g} {\gamma} \int_ {0} ^ {\rho_0} \rho f_-(\rho)\rd  \rho,
\end{align}
or, by the relations for volumes,
\begin{equation}\label{27}
\kappa -\frac{\varrho g}{2\gamma}h
= \frac{\sin \tilde\vartheta }{\rho_0} - \frac {\varrho g} {2 \pi \gamma \,\rho_0^2}
(V+V_s)
\end{equation}
So, for case in Fig.~3 the identity finds a slightly different form than others.


\section{Check of identities}
In order to provide the numerical tests of the identities obtained in previous section,
here a collection of numerical solutions is presented which covers
all the situations for which the identities are claimed to hold.
The outputs of numerical solutions, including the contact radius $\rho_0$,
apex height $h$ and curvature $\kappa$, and in case for curved substrates
the angle $\tilde\vartheta$, are presented in Tabs.~1-4, by which the direct
tests of identities are made possible. As illustrations, the plots
of all of the numerical solutions are presented in Figs.~4-7.

\begin{table}[h]{\scriptsize
\begin{center}
\begin{tabular}{c|ccc c  c }
Cont. angl.&  $\varrho$ & Bond & $\rho_0$ & $h$ & $\kappa$  \\
 & g/cm$^3$  & no. & cm  & cm &  cm$^{-1}$ \\
\hline
& & & &  &  \\
& 0 & 0 & 0.526 & 0.218 & 1.345  \\
$\vartheta=45^\circ$
& 0.5 & 1.51 & 0.546 & 0.194  & 0.990   \\
& 1.0 & 3.02 & 0.564 & 0.175  & 0.729 \\
& 1.5 & 4.52 & 0.581 & 0.159  & 0.536
\end{tabular}
\caption{{\small The geometrical values by the numerical solutions of Young-Laplace
relation for drops with acute contact angle, plotted in Fig.~4. For all:
$V=0.1~\mathrm{cm^3}$, $g=980~\mathrm{cm/s^2}$, $\gamma=70~\mathrm{dyn/cm}$.}}
\end{center}
}\end{table}

\begin{figure}[h]
\begin{center}
\includegraphics[width=0.6\columnwidth]{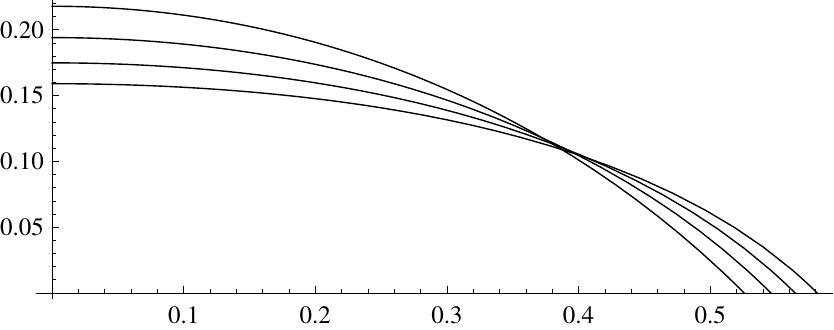}
\caption{\small The plots of numerical solutions of Young-Laplace relation for drops on flat
substrate with acute contact angle, given at Tab.~1 (scales: 1:1).}
\end{center}
\end{figure}

\begin{table}[h]{\scriptsize
\begin{center}
\begin{tabular}{c|ccc c  c  }
Cont. angl.&  $\varrho$ & Bond & $\rho_0$ & $h$ & $\kappa$  \\
 & g/cm$^3$  & no. & cm  & cm &  cm$^{-1}$  \\
\hline
& & & &  &  \\
& 0 & 0 & 0.208 & 0.501 & 3.404  \\
$\vartheta=135^\circ$
& 0.5 & 1.51 & 0.264 & 0.420 & 2.799   \\
& 1.0 & 3.02 & 0.295 & 0.375 & 2.332  \\
& 1.5 & 4.52 & 0.317 & 0.343 & 1.954
\end{tabular}
\caption{{\small The geometrical values by the numerical solutions of Young-Laplace
relation for drops with obtuse contact angle, plotted in Fig.~5. For all:
$V=0.1~\mathrm{cm^3}$, $g=980~\mathrm{cm/s^2}$, $\gamma=70~\mathrm{dyn/cm}$.}}
\end{center}
}\end{table}

\begin{figure}[h]
\begin{center}
\includegraphics[width=0.4\columnwidth]{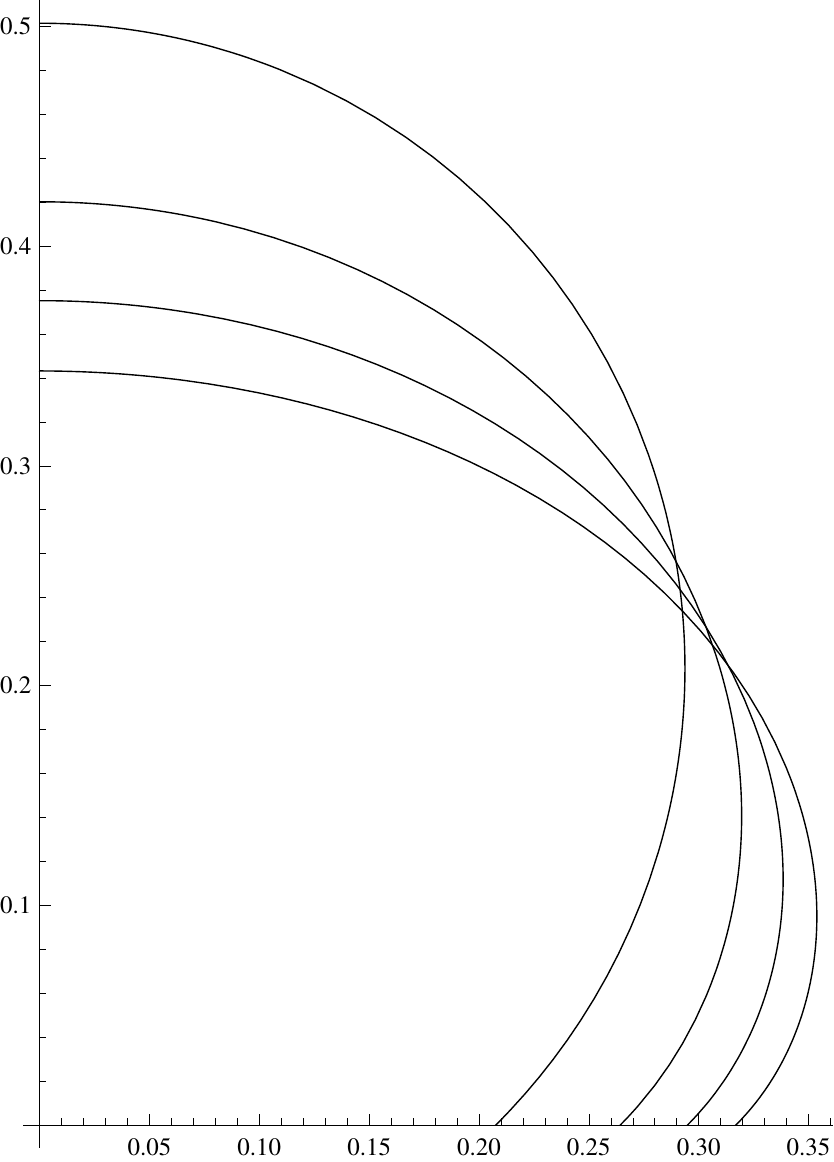}
\caption{\small The plots of numerical solutions of Young-Laplace relation for drops on flat
substrate with obtuse contact angle, given at Tab.~2 (scales: 1:1).}
\end{center}
\end{figure}

\begin{table}[h]{\scriptsize
\begin{center}
\begin{tabular}{c|ccc c c   c  }
\underline{Cont. angl.}&  $\varrho$ & Bond & $\rho_0$ & $h$ &  $\tilde\vartheta$ &   $\kappa$  \\
Subst. & g/cm$^3$  & no. & cm  & cm & deg.  & cm$^{-1}$  \\
\hline
& & & &  & &  \\
\underline{ $\vartheta=45^\circ$} &
 0 & 0 & 0.477 & 0.223 & 70.51 &   1.976  \\
Convex &
 0.5 & 1.51 & 0.517 & 0.175 & 72.32 &   1.417  \\
$s=-0.5\,\rho^2$ &
 1.0 &  3.02 & 0.562 & 0.129 & 73.33 &  0.967 \\
upp. Fig.~6 &
 1.5 & 4.52 & 0.631 & 0.073 & 77.23 &  0.577 \\
\hline
& & & &  & &  \\
\underline{$\vartheta=45^\circ$} &
 0 & 0.0 & 0.537 & 0.216 & 32.88 &  1.011 \\
Concave &
 1.0 & 3.02 & 0.561 & 0.188 & 32.35 &  0.565  \\
$s=0.2\,\rho^2$ &
 2.0 & 6.03 & 0.579 & 0.170 & 31.97 &  0.328  \\
mid. Fig~6 &
 3.0 & 9.05 & 0.592 & 0.158 & 31.67 &  0.200  \\
\hline
& & & &  & &  \\
\underline{$\vartheta=20^\circ$} &
 0 & 0 & 0.527 & 0.236 & $37.67$ & 1.160 \\
Concave &
 1.0 & 3.02 &0.512 & 0.258 & 36.94 & 0.751  \\
$s=1.5\,\rho^2$ &
 2.0 & 6.03 & 0.504 & 0.269 & 36.53 &  0.526  \\
low. Fig.~6 &
 4.0 & 12.1 & 0.495 & 0.282 & 36.06 &  0.289
\end{tabular}
\caption{{\small The geometrical values by the numerical solutions of Young-Laplace
relation for drops with acute contact angle, plotted in Fig.~6. For all:
$V=0.1~\mathrm{cm^3}$, $g=980~\mathrm{cm/s^2}$, $\gamma=70~\mathrm{dyn/cm}$.
For substrate $s(\rho)=\lambda\, \rho^2$, $V_s=2\pi\lambda\, \rho^4/4$.}}
\end{center}
}\end{table}

\begin{table}[h]{\scriptsize
\begin{center}
\begin{tabular}{c|ccc c c    c }
\underline{Cont. angl.}&  $\varrho$ & Bond & $\rho_0$ & $h$ &  $\tilde\vartheta$ &  $\kappa$   \\
Subst. & g/cm$^3$  & no. & cm  & cm & deg. & cm$^{-1}$  \\
\hline
& & & & & &  \\
\underline{ $\vartheta=135^\circ$} &
 0 & 0 & 0.169 & 0.514 & 144.6 & 3.438  \\
Convex &
 0.5 & 1.51 & 0.233 & 0.414 & 148.1 & 2.827  \\
$s=-0.5\,\rho^2$ &
1.0 & 3.02 & 0.268 & 0.359 & 150.0 & 2.346 \\
upp. Fig.~7 &
1.5 & 4.52 & 0.293 & 0.319 & 151.3 & 1.950 \\
\hline
& & & & & &  \\
\underline{$\vartheta=135^\circ$} &
 0 & 0 & 0.259 & 0.487 & 120.5 & 3.327 \\
Concave &
1.0 & 3.02 & 0.324 & 0.394 & 117.1 & 2.301  \\
$s=0.5\,\rho^2$ &
2.0 & 6.03 & 0.353 & 0.351 & 115.6 & 1.661  \\
low. Fig.~7 &
3.0 & 9.05 & 0.372 & 0.323 & 114.6 & 1.216
\end{tabular}
\caption{{\small The geometrical values by the numerical solutions of Young-Laplace
relation for drops with obtuse contact angle, plotted in Fig.~7. For all:
$V=0.1~\mathrm{cm^3}$, $g=980~\mathrm{cm/s^2}$, $\gamma=70~\mathrm{dyn/cm}$.
For substrate $s(\rho)=\lambda\, \rho^2$, $V_s=2\pi\lambda\, \rho^4/4$.}}
\end{center}
}\end{table}

\begin{figure}[h]
\begin{center}
\includegraphics[width=0.4\columnwidth]{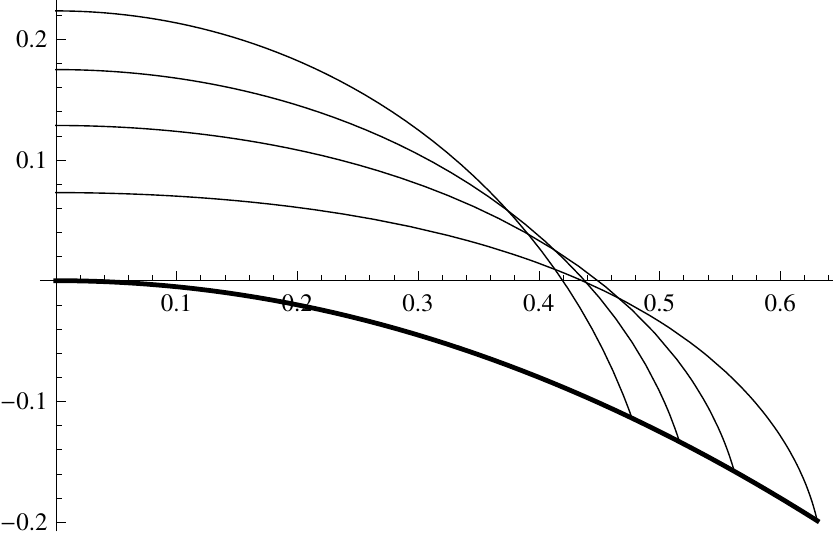}
\vskip 0.5cm
\includegraphics[width=0.4\columnwidth]{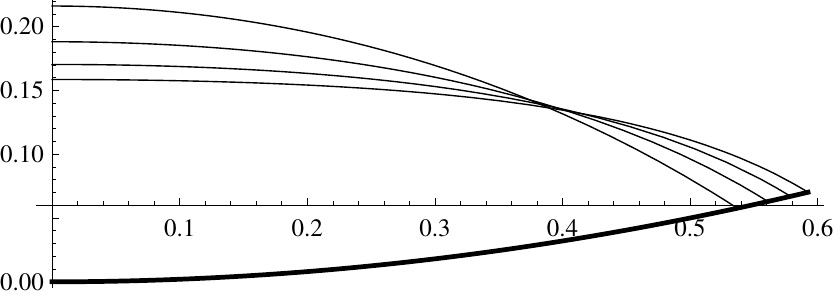}
\vskip 0.5cm
\includegraphics[width=0.4\columnwidth]{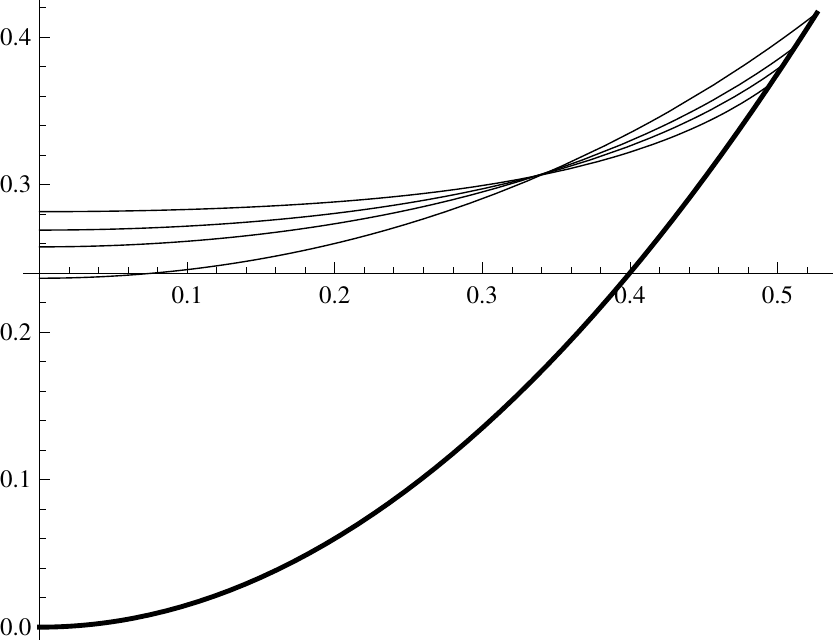}
\caption{\small The plots of numerical solutions of Young-Laplace relation for drops on
curved substrate with acute contact angle, given at Tab.~3 (scales: 1:1).}
\label{fig4}
\end{center}
\end{figure}

\begin{figure}[h]
\begin{center}
\includegraphics[width=0.3\columnwidth]{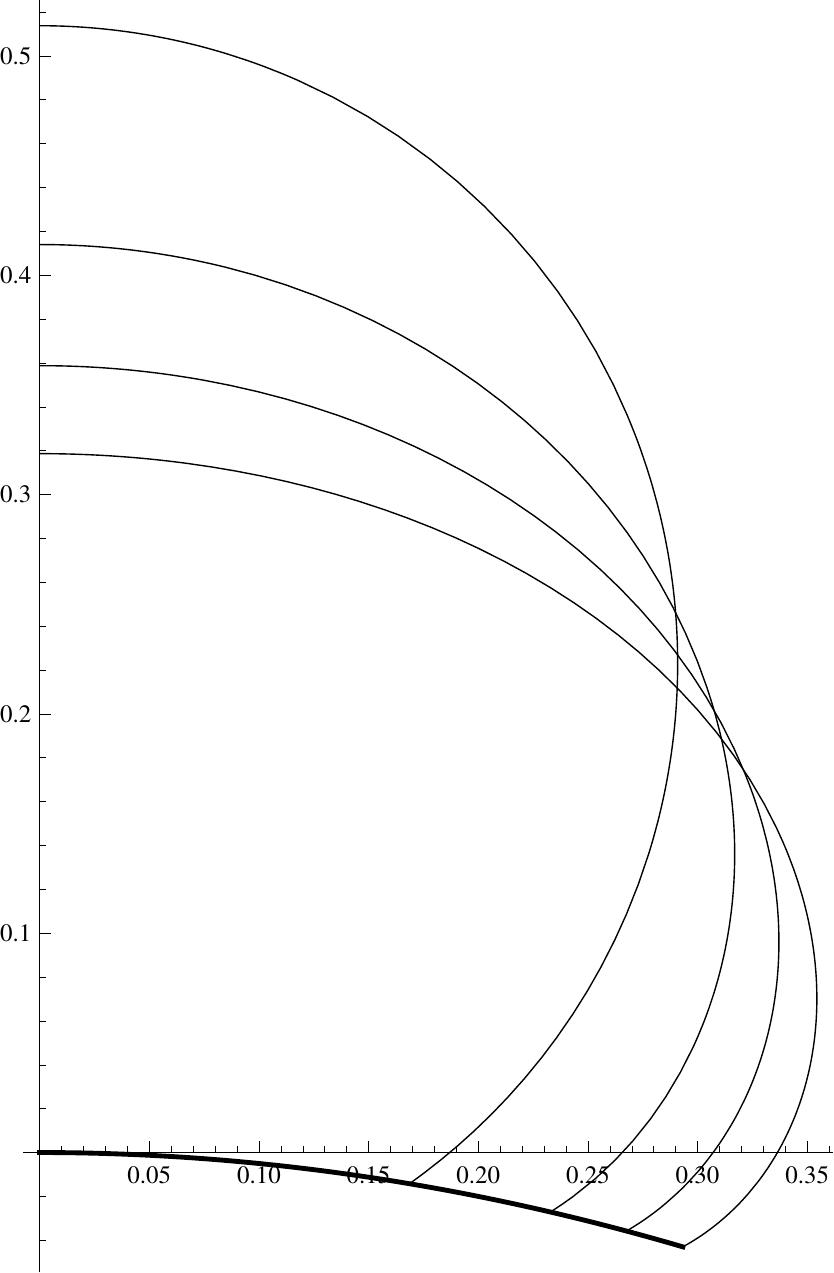}
\vskip 0.5cm
\includegraphics[width=0.3\columnwidth]{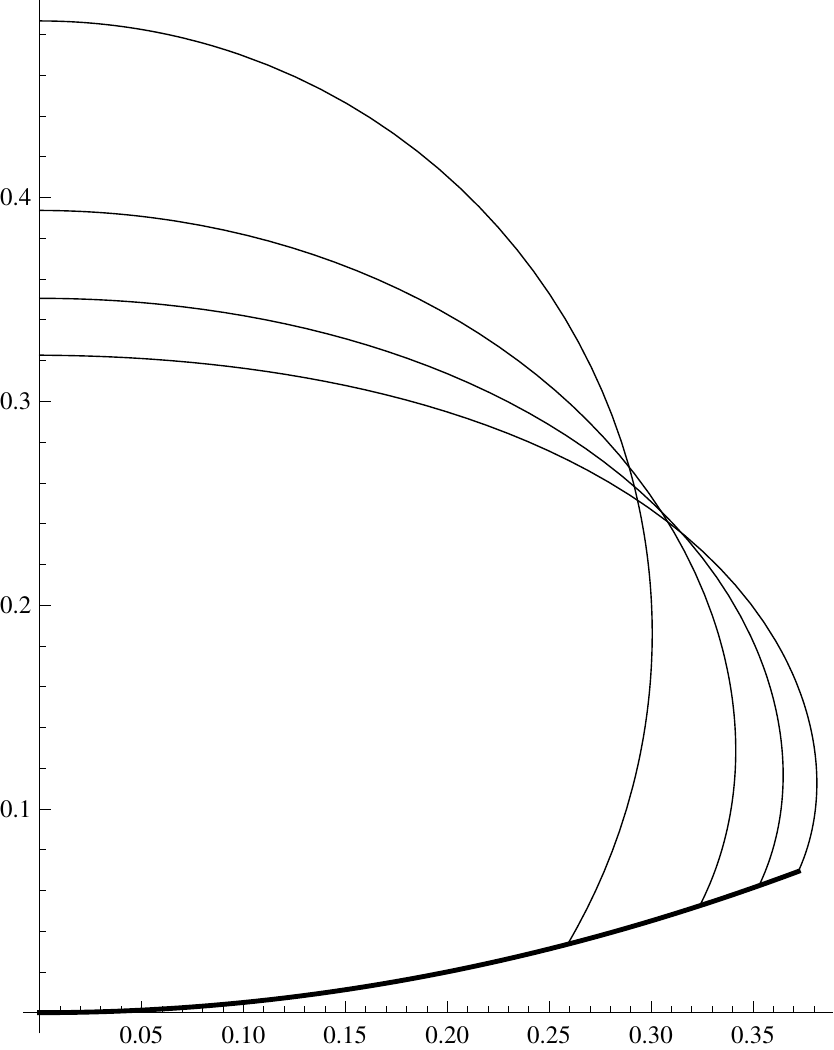}
\caption{\small The plots of numerical solutions of Young-Laplace relation for drops on
curved substrate with obtuse contact angle, given at Tab.~4 (scales: 1:1).}
\label{fig4}
\end{center}
\end{figure}


\vskip 0.5cm
\textbf{Acknowledgement}:
The authors are grateful to A. Aghamohammadi for helpful discussions.
The work by A.~H.~F. is supported by the Research Council of the Alzahra University.


\begin{thebibliography}{99}
\bibitem{adams} F. Bashforth and J.C. Adams, ``\textit{An attempt to test the theories of capillary attraction}", Cambridge University Press, Cambridge (1883).

\bibitem{staicop} D. N. Staicopolus, ``\textit{The computation of surface tension and of contact angle by the sessile-drop method}" (I and II), J. Colloid Interface Sci. \textbf{17} (1962) 439, \textit{ibid.} \textbf{18} (1963) 793.

\bibitem{padday} J. F. Padday, ``\textit{The profiles of axially symmetric menisci}", Phil. Trans. R. Soc. Lond. A \textbf{269} (1971) 265.

\bibitem{hartland} S. Hartland \& R. W. Hartley, ``\textit{Axisymmetric fluid-liquid interfaces}", Elsevier, Amsterdam (1976).

\bibitem{chester} A. K. Chesters, ``\textit{An analytical solution for the profile and volume of a small drop or bubble symmetrical about the vertical axis}", J. of Fluid Mech. \textbf{81} (1977) 609.

\bibitem{ehrlich} R. Ehrlich, ``\textit{An alternative method for computing contact angle from the dimensions of a small sessile drop}", J. Colloid Interface Sci. \textbf{28} (1968) 5.

\bibitem{smith} P. G. Smith \& T. G. M. van de Ven, ``\textit{Profiles of slightly deformed axisymmetric drops}", J. Colloid Interface Sci. \textbf{97} (1984) 1.

\bibitem{shan82} M. E. R. Shanahan, ``\textit{An approximate theory describing the profile of a sessile drop}", J. Chem. Soc., Faraday Trans. I \textbf{78} (1982) 2701.

\bibitem{shan84} M. E. R. Shanahan, ``\textit{Profile and contact angle of small sessile drops}", J. Chem. Soc., Faraday Trans. I \textbf{80} (1984) 37.

\bibitem{rienstra} S. W. Rienstra, ``\textit{The shape of a sessile drop for small and large surface tension}", J. Eng. Math. \textbf{24} (1990) 193.

\bibitem{ryley} D. J. Ryley and B. H. Khoshaim, ``\textit{A new method of determining the contact angle made by a sessile drop upon a horizontal surface (sessile drop contact angle)}", J. Colloid Interface Sci. \textbf{59} (1977) 243.

\bibitem{lehman} W. M. Robertson and G. W. Lehman, ``\textit{The shape of a
sessile drop}", J.  Appl. Phys. \textbf{39} (1968) 1994.

\bibitem{obrien} S. B. G. O'Brien, ``\textit{On the shape of small sessile
and pendant drops by singular perturbation techniques}", J. Fluid Mech. \textbf{233}
(1991) 519.

\bibitem{maze} C. Maze \& G. Burnet, ``\textit{A non-linear regression method for
calculating surface tension and contact angle
from the shape of a sessile drop}", Surface Sci. \textbf{13} (1969) 451.

\bibitem{neumann} Y. Rotenberg, L.~Boruvka \& A.~W.~Neumann, ``\textit{Determination of surface tension and contact angle from the shapes of axisymmetric fluid interfaces}", J. Colloid Interface Sci. \textbf{93} (1983) 169; P.~Cheng, D.~Li, L.~Boruvka, Y.~Rotenberg \& A.~W. Neumann, ``\textit{Automation of axisymmetric drop shape analysis for measurements of interfacial tensions and contact angles}", Colloids Surf. \textbf{43} (1990) 151.

\bibitem{kwok} D. Y. H. Kwok, ``\textit{Contact angles and surface energies}", Ph.D. Thesis, University of Toronto, page 32 (also available at http://www.mie.utoronto.ca/labs/last/kwok/drop.html).

\bibitem{graham} J. Graham-Eagle and S. Pennell, ``\textit{Contact angle calculations from the contact/maximum diameter of sessile drops}", Int. J. for Num. Meth. in Fluids \textbf{32} (2000) 851.

\bibitem{neum97} O. I. del Rio and A.~W.~Neumann, ``\textit{Axisymmetric Drop Shape Analysis: Computational Methods for the Measurement of Interfacial Properties from the Shape
and Dimensions of Pendant and Sessile Drops}", J. Colloid Interface Sci. \textbf{196} (1997)
136.

\bibitem{kwok91} E. Moy, P. Cheng, Z. Policova, S.~Treppo, D.~Kwok, D. R.~Mack,
P.~M.~Sherman, and A.~W.~Neumann, ``\textit{Measurement of contact angles from the maximum diameter of non-wetting drops by means of a modified axisymxnetric drop shape analysis}", Colloids Surf. \textbf{58} (1991) 215.

\bibitem{fathscr} A. H. Fatollahi, ``\textit{On the shape of a lightweight drop on a
horizontal plane}", Physica Scripta \textbf{85} (2012) 045401.

\end{thebibliography}
\end{document}